\documentclass[11pt]{article}

\usepackage{nsf}
\usepackage{times}
\usepackage{cite}
\usepackage{epsfig}
\usepackage{wrapfig}
\usepackage{amsmath}
\usepackage{amsfonts}
\usepackage{verbatim}
\usepackage{todonotes}
\usepackage[justification=centering,font=small,md]{caption}

\usepackage{subfigure}
\usepackage{booktabs}
\usepackage{tipa}
\usepackage{multirow}
\usepackage{tabularx}
\usepackage{lipsum}
\usepackage[english]{babel}
\usepackage{enumitem}
\usepackage{url}

\renewcommand{\refname}{whatever}

\date{\today }

\begin{document}

\bibliographystyle{abbrv}

\thispagestyle{empty}
\textbf{\huge Redes sociales, participaci\'on ciudadana y la hip\'otesis del \textit{slacktivismo}:  lecciones del caso de ``El Bronco''} 
\\\\
\vspace{-0.3pc}
\emph{{\bf Philip N. Howard}, Oxford University, {\bf Saiph Savage}, West Virginia University and Universidad Nacional Autonoma de Mexico (UNAM), {\bf Claudia Flores-Saviaga}, West Virginia University, {\bf Carlos Toxtli}, West Virginia University, {\bf Andr\'es Monroy-Hern\'andez}, Microsoft Research}
\vspace*{0.1in}\\
\vspace{-0.3pc}{{\bf Traducci\'on por:} Jos\'e Emilio Garc\'ia, Universidad Nacional Aut\'onoma de M\'exico (UNAM)}
\vspace{0.4pc}
\vspace{-0.3pc}

\begin{abstract}
?`El uso de las redes sociales tiene consecuencias positivas o negativas en la participaci\'on ciudadana? De acuerdo con la visi\'on c\'\i{}nica de la hip\'otesis del slacktivismo (el slacktivismo es un t\'ermino peyorativo tambi\'en conocido como ``activismo de sill\'on''), si los ciudadanos utilizan las redes sociales para el di\'alogo pol\'\i{}tico, dichas interacciones ser\'an fr\'\i{}volas y fugaces. La gran parte de los intentos por responder a esta pregunta incluyen datos de la opini\'on p\'ublica de los Estados Unidos, por lo que nosotros ofrecemos un estudio sobre un caso significativo de M\'exico, donde un candidato independiente utiliz\'o las redes sociales para comunicarse con el p\'ublico y rehuy\'o de los medios de comunicaci\'on tradicionales. Dicho candidato, conocido como ``El Bronco'', gan\'o la carrera por la gubernatura del estado al derrotar a los candidatos de los partidos tradicionales. Adem\'as, gener\'o una participaci\'on ciudadana que se ha mantenido m\'as all\'a del d\'\i{}a de las elecciones. En nuestra investigaci\'on analizamos m\'as de 750 mil mensajes, comentarios y respuestas durante m\'as de tres a\~nos de interacciones en la p\'agina p\'ublica de Facebook de ``El Bronco''. Examinamos la forma en que los ritmos de comunicaci\'on entre “El Bronco” y los usuarios se modificaron con el tiempo y demostramos que las redes sociales pueden utilizarse para dar cabida a una gran cantidad de interacciones ciudadanas sobre la vida p\'ublica m\'as all\'a de un acontecimiento pol\'\i{}tico particular.\footnote{Publicado en: \textit{Journal Of International Affairs}, Winter 2016, Vol. 70 Issue 1, p 55-73, \url{ISSN:0022-197X}}
\end{abstract}
\vspace{0.4pc}

\vspace{0.75pc}

\newpage

\setcounter{page}{1}
\setcounter{section}{3} 
\addcontentsline{toc}{section}{Project Description}



\textbf{\huge Redes sociales, participaci\'on ciudadana y la hip\'otesis del \textit{slacktivismo}:  lecciones del caso de ``El Bronco''
}


\vspace{-0.3pc}
\section{Introducci\'on}

Las redes sociales se han convertido en un elemento significativo de las campa\~nas pol\'\i{}ticas actuales. Los administradores de dichas campa\~nas, por ejemplo, las utilizan para obtener informaci\'on, mientras que los ciudadanos y diversas asociaciones civiles recurren a un amplio grupo de plataformas como Twitter, Facebook y reddit con el objetivo de hablar sobre pol\'\i{}tica e interactuar con grupos de la sociedad civil y l\'\i{}deres pol\'\i{}ticos \cite{karpf2012moveon}. Candidatos y partidos pol\'\i{}ticos tambi\'en utilizan las redes sociales para gestionar su imagen p\'ublica en sus interacciones con periodistas y el p\'ublico interesado \cite{gainous2014tweeting}. Si bien es cierto que existen diversos factores que determinan la adopci\'on de las redes sociales por parte de los actores pol\'\i{}ticos, cuando menos la gran mayor\'\i{}a se esfuerza por integrar las redes sociales a sus campa\~nas actualmente \cite{nah2012modeling}. Uno de los problemas radica en que, a pesar de los avances tecnol\'ogicos, la mayor parte de los grupos pol\'\i{}ticos y activistas a\'un desconocen el mejor m\'etodo para movilizar a la poblaci\'on \cite{lovejoy2012information}. La mayor dificultad se debe al hecho de que las decisiones de los ciudadanos sobre su participaci\'on en una causa pol\'\i{}tica dependen de su percepci\'on de los esfuerzos de un candidato u organizaci\'on \cite{savage2016botivist}. Sin embargo, las normas que rigen las plataformas de redes sociales se encuentran en cambio constante. Como resultado, las personas modifican continuamente la manera en que interpretan el contenido de las redes sociales y los esfuerzos en l\'\i{}nea de los dem\'as. De esta forma, para los grupos pol\'\i{}ticos podr\'\i{}a resultar en extremo dif\'\i{}cil y costoso en t\'erminos de tiempo mantenerse actualizados con las nuevas tecnolog\'\i{}as y predecir el resultado de compartir ciertos tipos de contenidos. Esta complejidad ha obligado a varias organizaciones pol\'\i{}ticas y candidatos a limitar el tiempo que utilizan las redes sociales y a delimitar el personal que tiene acceso a ellas \cite{obar2014canadian}. Por lo general los pol\'\i{}ticos prefieren tener una persona a cargo de la estrategia de su partido en redes sociales e incluso a cargo de sus cuentas personales. No obstante, aun cuando la persona encargada ha encontrado la forma de movilizar a los ciudadanos, no resulta sencillo transmitir este conocimiento a los dem\'as. Por lo tanto, la mayor\'\i{}a de los pol\'\i{}ticos suelen ser muy precavidos con respecto a la cantidad tiempo que utilizan las redes sociales y al grado en que interact\'uan con sus p\'ublicos en l\'\i{}nea  \cite{savage2015participatory}. Esta situaci\'on, por su parte, ha obstaculizado y limitado nuestro entendimiento sobre las estrategias pol\'\i{}ticas m\'as adecuadas para movilizar a los ciudadanos y vincularse con ellos.

Nuestra idea de las redes sociales y de las elecciones est\'a delimitada por el hecho de que la mayor parte de la investigaci\'on est\'a enfocada en Estados Unidos. En esta democracia relativamente avanzada, las redes sociales se han convertido en una herramienta para algunas formas de participaci\'on ciudadana y de expresi\'on pol\'\i{}tica. Sin embargo, en las democracias avanzadas esto s\'olo parece haber tenido como resultado modestas expresiones de activismo, como firmar peticiones o compartir contenido pol\'\i{}tico de grupos afines en redes formadas por familiares y amigos \cite{gibson2000social}. Por lo tanto, la investigaci\'on sobre las redes sociales y el involucramiento de la sociedad civil en la pol\'\i{}tica estadounidense con frecuencia ha intentado comprobar, directa o indirectamente, la “hip\'otesis del slacktivismo”. Por \'esta entendemos la suposici\'on de que al aumentar el uso de Internet o de las redes sociales, el involucramiento de la sociedad civil se reduce.

La idea de que el uso de las redes sociales tiene consecuencias predominante negativas para la vida p\'ublica se basa en primera instancia en la evidencia de que la mayor parte del contenido compartido en las redes sociales rara vez se relaciona con la pol\'\i{}tica, e incluso cuando lo hace no suele consistir m\'as que en breves mensajes en breves conversaciones con gente irritable \cite{katz2013social}. Con muy poca frecuencia los ciudadanos utilizan las redes sociales para tener di\'alogos pol\'\i{}ticos sustanciales, los cuales suelen ser descorteses y polarizados. En general, las conversaciones en l\'\i{}nea sobre pol\'\i{}tica son relativamente raras en comparaci\'on con otro tipo de actividades que las personas llevan a cabo diariamente en Internet \cite{massanari2011information}. Adem\'as, cuando \'estas ocurren durante acontecimientos pol\'\i{}ticos significativos —como debates entre candidatos— los usuarios de las redes sociales recurren a las plataformas digitales para informarse sobre pol\'\i{}tica, pero tienden a adquirir conocimiento que s\'olo favorece a su candidato preferido \cite{boulianne2015social}. Investigaciones recientes han encontrado que, si bien diversas organizaciones activistas creen estar creando comunidades m\'as fuertes y propiciando el di\'alogo con su p\'ublico a trav\'es de su contenido en redes sociales, en realidad esto rara vez se traduce en una movilizaci\'on significativa relacionada con eventos p\'ublicos, activismo del consumidor o cabildeo comunitario \cite{guo2013tweeting,lovejoy2012information}.

Investigadores han demostrado que el uso de las redes sociales provoca que las personas conviertan sus redes sociales en “burbujas de filtro”, las cuales disminuyen las posibilidades de ser expuestas a ideas nuevas o desafiantes. En otras palabras, las redes sociales permiten crear redes de homofilia \cite{pariser2011filter}. Por ejemplo, grandes cantidades de datos tomados de Twitter han sido utilizados para clasificar a los usuarios por su afiliaci\'on partidaria y homofolia en Estados Unidos. Los resultados indican que los dem\'ocratas promedio tienden a ser m\'as homof\'\i{}licos que los republicanos promedio, excepto en los casos en que los usuarios clasificados como republicanos siguen a los principales l\'\i{}deres de su partido \cite{colleoni2014echo}. En \'ultima instancia, es posible que los debates p\'ublicos a trav\'es de las redes sociales no hagan mucho m\'as que promover una participaci\'on ef\'\i{}mera carente de consecuencias fuera de las redes sociales \cite{christensen2011political,rotman2011slacktivism}. En los casos en que las acciones en redes sociales tienen consecuencias fuera de \'estas, suelen ser el mismo tipo de acciones de baja calidad y de gran volumen que los grupos pol\'\i{}ticos y de defensa han utilizado durante largo tiempo para adquirir notoriedad y aparecer en los titulares \cite{karpf2010online}.

En las primeras etapas del debate sobre el slacktivismo, tan s\'olo parec\'\i{}an existir algunos casos muy espec\'\i{}ficos provenientes de Internet en que se hab\'\i{}a logrado movilizar con \'exito a la poblaci\'on y, al mismo tiempo, alcanzar metas de pol\'\i{}tica p\'ublica. Adem\'as, dichos casos sol\'\i{}an implicar una estrecha gama de problemas relacionados con el acceso a la tecnolog\'\i{}a \cite{benkler2015social,faris2015score,freedman2016strategies, dubois2012fifth,kreiss2016prototype,nielsen2012ground}. En a\~nos m\'as recientes, sin embargo, la distinci\'on entre las acciones pol\'\i{}ticas en l\'\i{}nea y aqu\'ellas fuera de Internet se ha evaporado, de manera que los candidatos actuales necesitan conocer a fondo diversas plataformas tecnol\'ogicas, al tiempo que las campa\~nas pol\'\i{}ticas invierten recursos significativos en el an\'alisis de datos. Existen m\'ultiples ejemplos de movimientos sociales tradicionales que han obtenido victorias impresionantes a trav\'es del uso efectivo de las redes sociales, as\'\i{} como de nuevos movimientos que se originaron en Internet y se han convertido en actores estables de la sociedad civil \cite{beyer2014expect}. Otro elemento que complica esta situaci\'on es el creciente problema del control algor\'\i{}tmico de los mensajes en las redes sociales: los programas automatizados pueden ser utilizados tanto para movilizar a los ciudadanos como para desalentar su participaci\'on \cite{savage2015participatory,woolley2016automation}.

El argumento en contra de la hip\'otesis del slacktivismo est\'a basado en la idea de que la participaci\'on pol\'\i{}tica en las redes sociales es siempre un a\~nadido —y no un sustituto— de aquello que los ciudadanos realizar\'\i{}an normalmente en su vida pol\'\i{}tica \cite{christensen2011political}. Existen conversaciones p\'ublicas significativas que se dan a trav\'es de las redes sociales y que se intensifican enormemente durante sucesos pol\'\i{}ticos importantes. Por ejemplo, investigaciones han encontrado que el uso de las redes sociales ayuda a las personas a construir su identidad pol\'\i{}tica, as\'\i{} como su consciencia comunitaria, lo que incluso da lugar a contribuciones financieras a importantes grupos de la sociedad civil \cite{lee2013does}. De hecho, las redes sociales —as\'\i{} como otros tipos de comunicaci\'on basados en Internet— tienden a complementar nuestro consumo de informaci\'on sobre pol\'\i{}tica, elecciones y pol\'\i{}ticas p\'ublicas, adem\'as de permitir a las personas, por as\'\i{} decirlo, llevar una dieta de informaci\'on m\'as omn\'\i{}vora. Estos “omn\'\i{}voros pol\'\i{}ticos” a\'un recurren a los principales medios de difusi\'on para obtener informaci\'on, sin embargo, dependen de Internet para sus interacciones sobre pol\'\i{}tica \cite{massanari2011information}.

Existe evidencia de que el uso de las redes sociales por parte de adolescentes —junto con la deseo de participar y el consumo de noticias televisivas— crea un c\'\i{}rculo virtuoso de participaci\'on ciudadana \cite{kruikemeier2016news}. En consecuencia, la mayor parte de la investigaci\'on se ha enfocado principalmente en encuestas a peque\~na escala o en estudios basados en entrevistas. Adem\'as, ha resultado dif\'\i{}cil medir los niveles de participaci\'on ciudadana en las plataformas de redes sociales, sobre todo porque usualmente no existe un sitio central de dichas redes que utilicen tanto ciudadanos como pol\'\i{}ticos. El estar expuesto al activismo en l\'\i{}nea podr\'\i{}a influenciar las decisiones individuales alrededor de acciones ciudadana posteriores, como firmar una petici\'on o donar a la caridad; sin embargo, no es claro que el uso de las redes sociales para hablar sobre pol\'\i{}tica tenga como resultado votantes m\'as sofisticados o una mayor probabilidad de participaci\'on electoral.

Si bien los investigadores debaten sobre la hip\'otesis del slacktivismo en el contexto de Estados Unidos y las democracias avanzadas, existen buenas razones para creer que la relaci\'on entre el uso de las redes sociales y la participaci\'on ciudadana es diferente en contextos internacionales. La primera raz\'on resulta bastante clara para los investigadores de estudios internacionales: hay tal variedad de tipos de reg\'\i{}menes e instituciones pol\'\i{}ticas en el mundo que no debemos esperar que los resultados de Estados Unidos tengan validez en muchos otros contextos. La segunda raz\'on proviene de una observaci\'on m\'as espec\'\i{}fica de la investigaci\'on comparativa de sistemas de medios: fuera de Estados Unidos, se producen, consumen y regulan las noticias y la informaci\'on pol\'\i{}tica de maneras muy diferentes \cite{wessler2016global}.

En primer lugar, para la mayor parte del mundo Facebook abarca la principal experiencia en Internet, en efecto, una cantidad significativa del tiempo que muchos usuarios pasan en l\'\i{}nea lo dedican a esta plataforma \cite{stewart2016facebook}. El Internet que activistas, ciudadanos y votantes utilizan para consumir informaci\'on pol\'\i{}tica en Estados Unidos y en varias democracias avanzadas es experimentado a trav\'es de un navegador en una computadora personal, y cada vez m\'as, en un tel\'efono inteligente. En regiones del mundo donde los planes de datos son m\'as costosos y donde el ancho de banda es m\'as err\'atico, las personas utilizan en su lugar SMS (mensajes de texto) o las aplicaciones que vienen preinstaladas en los tel\'efonos celulares m\'as econ\'omicos. Como resultado, Facebook se ha convertido en sin\'onimo de Internet para la mayor\'\i{}a de personas en el mundo.

Sin embargo, la inmensa mayor\'\i{}a de las investigaciones sobre el slacktivismo se ha realizado en Estados Unidos, y estudios de caso internacionales sobre el tema demuestran que la relaci\'on entre la difusi\'on de las redes sociales y la participaci\'on ciudadana es compleja. Algunos investigadores han demostrado que los l\'\i{}deres pol\'\i{}ticos raramente utilizan las plataformas m\'as interactivas por miedo a perder el control del contenido que producen y de los mensajes que elaboran \cite{stromer2014presidential,howard2003digitizing}. Morozov ha argumentado que diversas plataformas de medios digitales son incapaces de mantener la atenci\'on de las personas que tan s\'olo ofrecen unos clics de apoyo a trav\'es de peticiones en l\'\i{}nea pero cuentan con poca energ\'\i{}a para los tipos de participaci\'on pol\'\i{}tica que toman tiempo o que implican un riesgo personal \cite{morozov2014save}. Morozov se basa en varios ejemplos internacionales de movimientos que pudieron haber fracasado a causa de su dependencia en las tecnolog\'\i{}as de la informaci\'on; su an\'alisis, sin embargo, no es el m\'as sistem\'atico. 

La bibliograf\'\i{}a sobre comunicaci\'on internacional es extensa, pero hay lecciones que tomar en cuenta de investigadores que abordan el estudio de la comunicaci\'on pol\'\i{}tica contempor\'anea. Un estudio detallado de los reg\'\i{}menes autoritarios, desde Azerbaiy\'an hasta China, revel\'o que las redes sociales pueden ser un medio significativo para dialogar sobre pol\'\i{}tica. Sin embargo, esta posibilidad depende de los intereses de censura y de control de la informaci\'on por parte de las \'elites gobernantes \cite{king2016chinese,king2013censorship,pearce2015affordances}. Bailard, por su parte, ha demostrado a trav\'es de un experimento de campo que el uso de los medios digitales durante las elecciones en Tanzania y Bosnia tuvo como resultado un incremento en los niveles de participaci\'on ciudadana, un efecto insignificante en la participaci\'on electoral, y un aumento en la tasa de cinismo electoral \cite{anderson2016democracy,bailard2012field}. M\'as a\'un, el uso de las redes sociales es una parte significativa de la explicaci\'on causal de la forma y el car\'acter de un n\'umero creciente de importantes protestas p\'ublicas, incluyendo aqu\'ellas en Chile y Hong Kong, en las que el uso de Facebook y Twitter para enterarse de las noticias result\'o ser un destacado predictor de la participaci\'on en las protestas, incluso teniendo en cuenta factores constantes como los valores postmaterialistas y la ideolog\'\i{}a pol\'\i{}tica \cite{valenzuela2012social,lee2015social}. Un estudio a largo plazo sobre la confianza en las instituciones en siete pa\'\i{}ses de Asia revel\'o que las redes sociales tienen un papel particularmente importante en el aumento de la participaci\'on ciudadana en pa\'\i{}ses donde otros medios son escasos \cite{lee2016digital}. Uno de los primeros estudios sobre cuatro democracias avanzadas revel\'o asimismo que los temores de que Internet tuviera un amplio efecto negativo sobre el capital social y la participaci\'on pol\'\i{}tica eran infundados \cite{gibson2000social}.

El florecimiento de las protestas en favor de la democracia tambi\'en podr\'\i{}a socavar la hip\'otesis del slacktivismo. Los movimientos sociales son fen\'omenos causalmente complejos; sin embargo, tanto activistas como l\'\i{}deres de protestas afirman que las redes sociales fueron esenciales para la organizaci\'on de las manifestaciones. Los l\'\i{}deres del gobierno, por su parte, intentan responder en las redes sociales o censurarlas, pues juzgan que son parte integral de la protesta. A su vez, los periodistas informan sobre la din\'amica particular del uso de las redes sociales en sus reportajes, mientras que a los investigadores, en retrospectiva, les resulta dif\'\i{}cil desarrollar una narrativa anal\'\i{}tica sobre los sucesos sin abordar el papel de las redes sociales \cite{monroy2015narcotweets}.

\textbf{El caso de ``El Bronco''}

Encontrar casos internacionales que nos ayuden a comprobar la hip\'otesis del \textit{slacktivismo} es un desaf\'\i{}o, ya que existen pocos l\'\i{}deres pol\'\i{}ticos —ya sea en reg\'\i{}menes autoritarios o democr\'aticos— cuyas carreras se hayan construido a trav\'es de un uso experimentado de las redes sociales. Las pruebas que tenemos sobre las redes sociales y el \textit{slacktivismo} tambi\'en se encuentran limitadas por el hecho de que en cada uno de los sistemas de medios que abordamos, las redes sociales son tan s\'olo un peque\~na parte de la estrategia de comunicaci\'on de los l\'\i{}deres pol\'\i{}ticos. Hasta hace poco, las campa\~nas por medio de las redes sociales consist\'\i{}an principalmente en entrar en contacto con periodistas y encargados de pol\'\i{}ticas p\'ublicas, m\'as que con la poblaci\'on general \cite{kreiss2014seizing}. La mayor\'\i{}a de los sistemas de medios cuentan con herramientas dominantes a trav\'es de las cuales los ciudadanos obtienen noticias e informaci\'on sobre pol\'\i{}tica, y que por lo general son la televisi\'on, la radio o los peri\'odicos. El Internet —y en particular las redes sociales— provee una fuente secundaria de medios de comunicaci\'on, aunque curiosamente, se trata del medio secundario m\'as popular de entre todas las posibilidades \cite{massanari2011information}. En otras palabras, los ciudadanos a menudo obtienen la mayor parte de sus noticias e informaci\'on a trav\'es de la televisi\'on, la radio o el peri\'odico, y despu\'es revisan las fuentes, le preguntan a sus amigos y familiares o investigan en Internet.

No obstante, esta situaci\'on puede estar cambiando. En 2015, Jaime Rodr\'\i{}guez Calder\'on, un candidato independiente mexicano tambi\'en conocido como “El Bronco”, gan\'o la elecci\'on en el estado de Nuevo Le\'on basado en una campa\~na que trat\'o a los medios de comunicaci\'on tradicionales con desd\'en y, en cambio, particip\'o de manera activa con electorado a trav\'es de Facebook y Twitter \cite{broncoMex}. A diferencia de la mayor\'\i{}a de los candidatos, “El Bronco” no contrat\'o anuncios de televisi\'on. Incluso los anuncios en las calles —muy comunes en las elecciones en M\'exico— fueron hechos por simpatizantes de la propia comunidad y no por una organizaci\'on de campa\~na centralizada. En este art\'\i{}culo, analizamos la totalidad de las interacciones en l\'\i{}nea entre los ciudadanos y “El Bronco”, quien principalmente a trav\'es de Facebook fue capaz de postularse como candidato independiente y derrotar a su opositor m\'as cercano por casi 25 puntos porcentuales \cite{broncoMex2}.

Nuevo Le\'on es un estado del norte de M\'exico con alrededor de cinco millones de habitantes; ocupa el segundo lugar en el pa\'\i{}s tanto en los indicadores de desarrollo como en el \'\i{}ndice de penetraci\'on de Internet. Cerca del 60 por ciento de los hogares en Nuevo Le\'on tienen acceso a este medio, cifra comparable a estados como Mississippi en Estados Unidos \cite{broncoMex3,broncoMex4,rainie2014census}.

El caso de “El Bronco” resulta valioso precisamente porque es un caso extremo del uso de las redes sociales durante una elecci\'on competida. Adem\'as, destaca el notable resultado de un candidato independiente que gana una elecci\'on a trav\'es de la participaci\'on dedicada con los votantes en redes sociales. Esta importante elecci\'on regional ofrece una oportunidad \'unica para responder a una significativa pregunta de investigaci\'on: ?`cu\'al es la relevancia del uso de las redes sociales en la participaci\'on ciudadana durante la temporada de campa\~nas electorales y m\'as all\'a del d\'\i{}a de las elecciones?

\section{Colecci\'on de Datos y An\'alisis}
\begin{figure}
\centering
  \includegraphics[width=1.0\columnwidth]{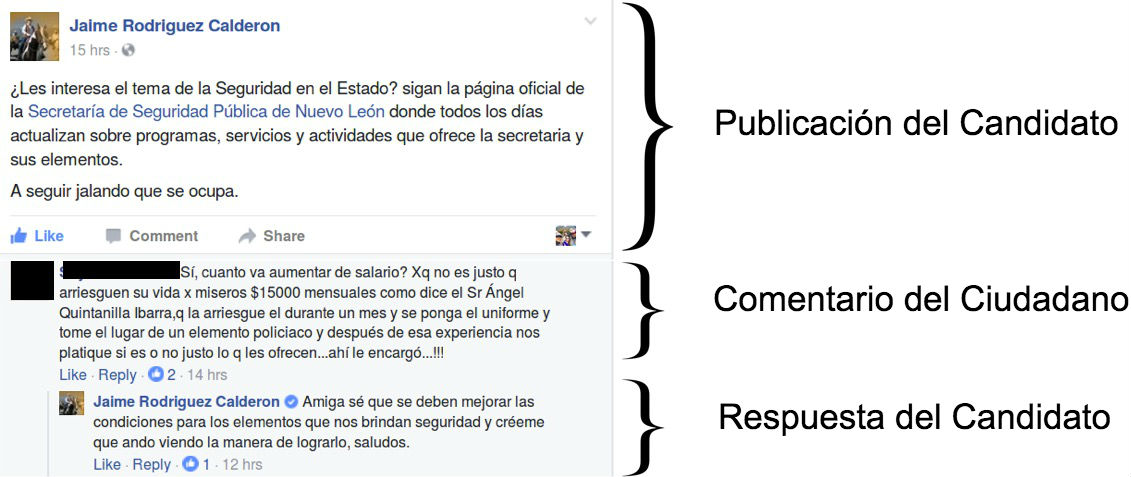}
  \caption{Conversaci\'on iniciada por Jaime Rodriguez Calderon. }\label{fig:postFacebook}
\end{figure}

Para estudiar los patrones a largo plazo de la participaci\'on pol\'\i{}tica en las redes sociales, recopilamos tres a\~nos de publicaciones, comentarios y respuestas de la p\'agina de seguidores de Facebook de Jaime Heliodoro Rodr\'\i{}guez Calder\'on \footnote{\url{https://www.facebook.com/JaimeRodriguezElBronco}}. Por medio de la interfaz de programaci\'on de aplicaciones (API por sus siglas en ingl\'es) de la plataforma, recopilamos 5 mil 708 publicaciones del candidato y 71 mil 446 de los usuarios; 20 mil 45 comentarios del candidato y 458 mil 544 de los usuarios; y 31 mil 527 respuestas a los comentarios por parte del candidato, as\'\i{} como 171 mil 577 de los usuarios. Este m\'etodo recopil\'o toda la actividad en la p\'agina entre noviembre de 2012 —fecha en que se cre\'o la presencia en las redes sociales de “El Bronco”— y abril de 2016 —casi un a\~no despu\'es de su toma de posesi\'on.  

Comenzamos nuestro an\'alisis con la recopilaci\'on de varios tipos de datos, provenientes de numerosas personas y generados en diversos tipos de dispositivos. Creamos una serie de bots conectados a Internet para que realizaran consultas a la API de Facebook y recopilaran datos sobre publicaciones, im\'agenes, comentarios y reacciones del tipo “me gusta” en la p\'agina de seguidores de “El Bronco”. Utilizamos MongoDB —una base de datos no relacional— para que nos ayudara a administrar el proceso de almacenaje de la gran cantidad de datos generados por nuestras consultas, y para compaginar los cambios en la forma en que Facebook organiza los datos sobre las publicaciones y los dispositivos de los usuarios. 

Dichos datos registran conversaciones iniciadas ya sea por el candidato o por un usuario de Facebook y consisten en las publicaciones en la p\'agina de “El Bronco”, los comentarios de los usuarios y las respuestas posteriores. Para ayudar a explicar la estructura de estos datos, la Gr\'afica 1 presenta la organizaci\'on de una conversaci\'on t\'\i{}pica iniciada por el candidato. \'este realiza un comentario que provoca otro por parte de un ciudadano, tras lo cual el candidato puede responder a dicho comentario. Esta secuencia puede desarrollarse a lo largo de una extensa serie de respuestas. La Gr\'afica 2 presenta la organizaci\'on de una conversaci\'on sobre pol\'\i{}tica iniciada por un ciudadano. \'este realiza una publicaci\'on en la p\'agina y posiblemente suscita la participaci\'on cuando el candidato comenta dicha publicaci\'on. El ciudadano puede responder y sostener un largo intercambio con el candidato. Las publicaciones, los comentarios y las respuestas pueden ser vistas por cualquier usuario, quienes tambi\'en pueden realizar sus propias publicaciones. 
\begin{figure}
\centering
  \includegraphics[width=.6\columnwidth]{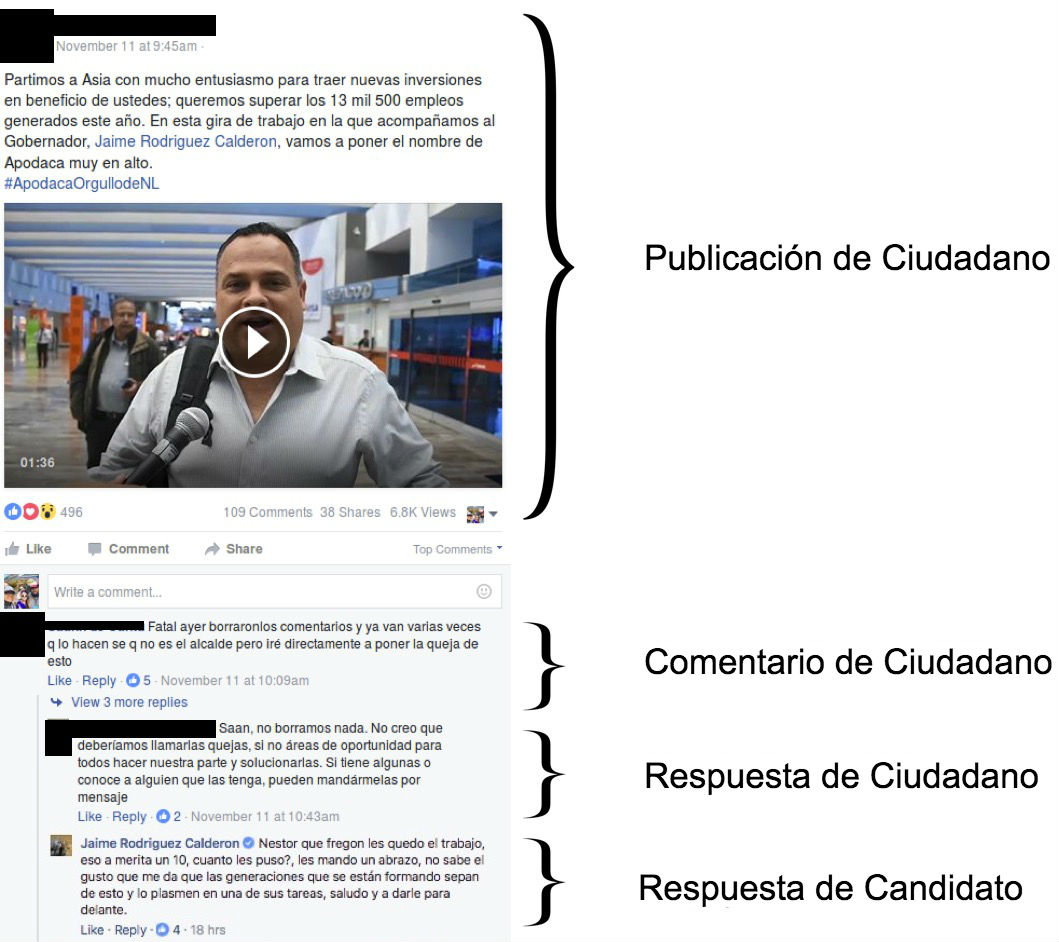}
  \caption{Conversaci\'on iniciada por un ciudadano. }\label{fig:postFacebook}
\end{figure}

Con el objetivo de estudiar los patrones de comunicaci\'on del candidato y los ciudadanos, graficamos las medias m\'oviles de la cantidad de publicaciones, comentarios y respuestas. Utilizamos una media m\'ovil de 15 d\'\i{}as que nos permitiera mostrar una tendencia uniforme durante el periodo de tres a\~nos y que fuera m\'as f\'acil interpretar.  

“El Bronco” se volvi\'o un pol\'\i{}tico independiente al dejar al Partido Revolucionario Institucional (PRI) y en noviembre de 2012, cre\'o una p\'agina de Facebook para elaborar su nuevo perfil pol\'\i{}tico. En ese tiempo, “El Bronco” afirm\'o que su ruptura con el PRI y su desaprobaci\'on de los principales medios de difusi\'on eran por s\'\i{} mismos importantes caracter\'\i{}sticas de su nuevo identidad p\'ublica. Durante gran parte de 2013, el tr\'afico en su plataforma de redes sociales fue limitado; sin embargo, el a\~no siguiente, su firme postura en contra del crimen organizado y la corrupci\'on del gobierno encontr\'o un p\'ublico en aumento. M\'as a\'un, ganaba popularidad varios meses antes de la elecci\'on, y para junio del mismo a\~no, hac\'\i{}a numerosas publicaciones, comentarios y respuestas todos los d\'\i{}as —un ritmo de participaci\'on en las redes sociales que pudo mantener durante los dos a\~nos siguientes—. 
\begin{figure}
\centering
  \includegraphics[width=.8\columnwidth]{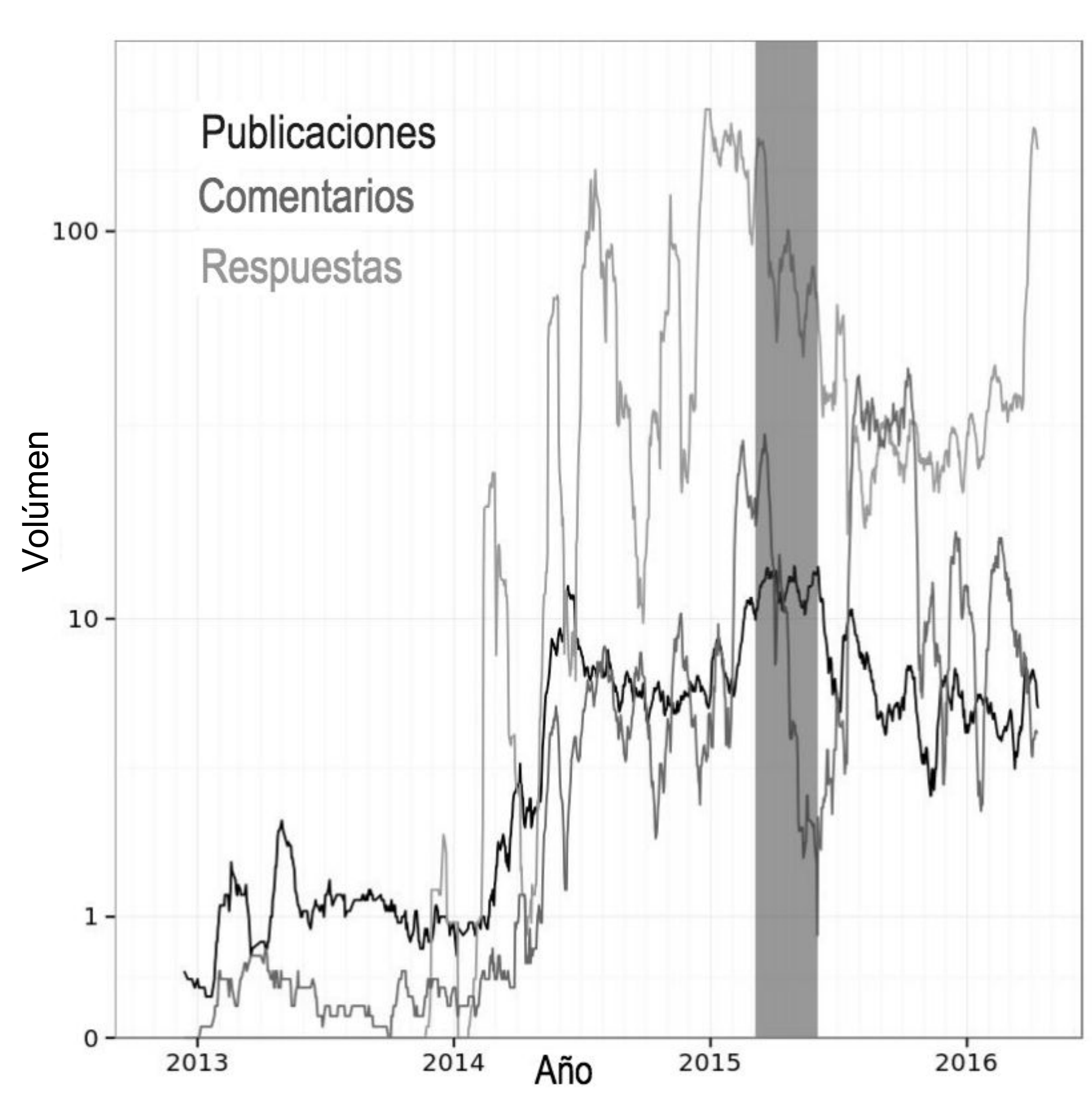}
  \caption{Ritmo de contenido generado por el candidato }\label{fig:postFacebook}
\end{figure}

\begin{figure}
\centering
  \includegraphics[width=.8\columnwidth]{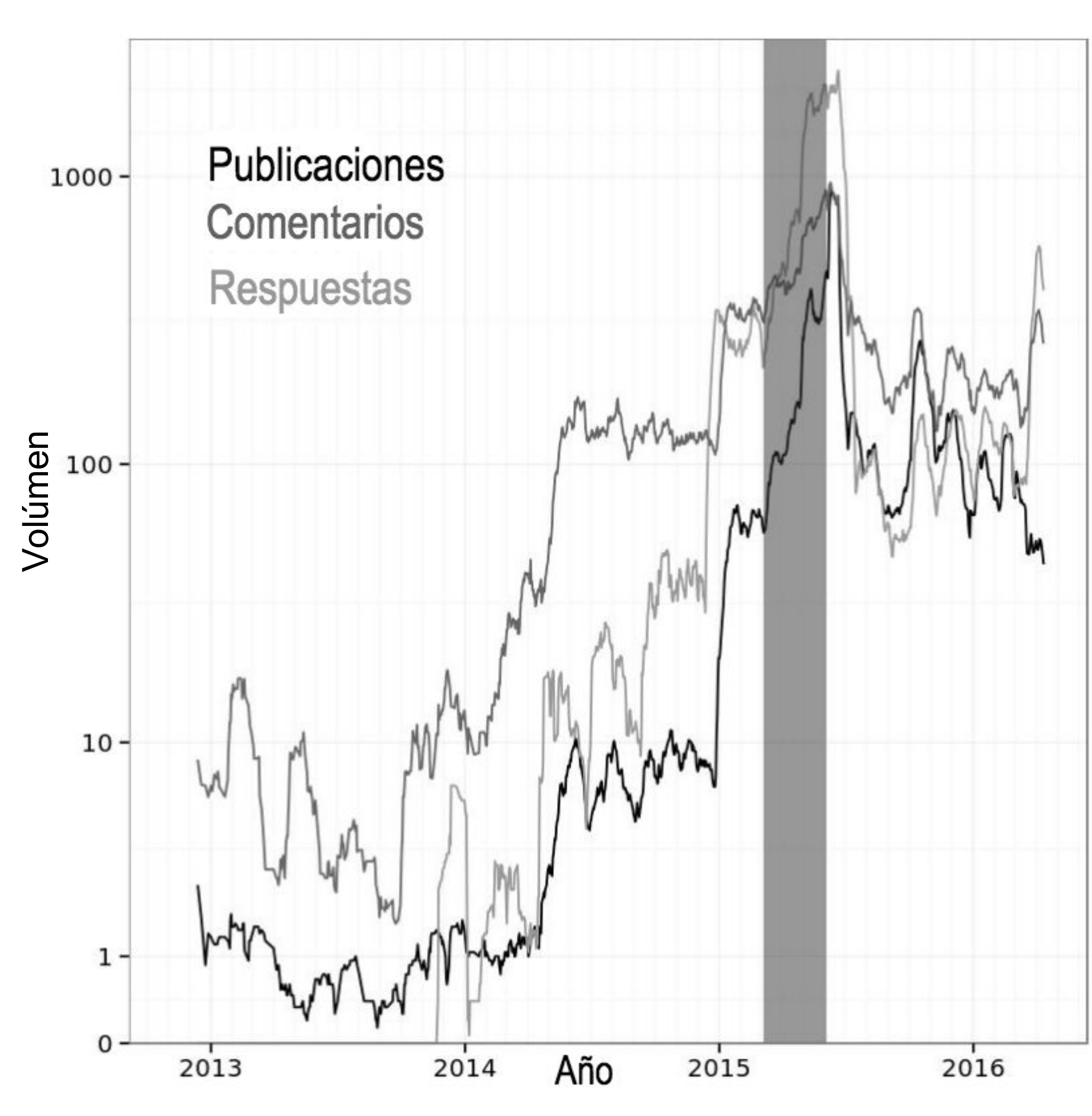}
  \caption{Ritmo de contenido generado por los ciudadanos }\label{fig:postFacebook}
\end{figure}

Las Gr\'aficas 3 y 4 representan la media m\'ovil del n\'umero de publicaciones, comentarios y respuestas durante el todo el periodo. A pesar de que la fase oficial de campa\~na para gobernador del estado s\'olo dur\'o tres meses, resulta evidente que la participaci\'on ciudadana en la p\'agina de Facebook despunt\'o mucho antes. De hecho, la narrativa del an\'alisis de estos datos deber\'\i{}a tener tres cap\'\i{}tulos: la etapa de precampa\~na, de noviembre de 2011 a febrero de 2015; el periodo oficial de campa\~na, de marzo a mayo de 2015; y el tiempo como gobernador, de junio del mismo a\~no hasta la fecha. En estas gr\'aficas muchos de los picos en la discusi\'on pol\'\i{}tica en las redes sociales tuvieron lugar durante momentos significativos de la pol\'\i{}tica nacional e internacional, mientras que varios de los valles coinciden con los d\'\i{}as feriados. La franja gris en las gr\'aficas indica el comienzo y el final de la campa\~na oficial para gobernador, que se desarroll\'o del 6 de marzo al 3 de junio de 2015.

La Gr\'afica 4 revela que el ritmo de la conversaci\'on p\'ublica, tanto en torno a la pol\'\i{}tica como al candidato en particular, se increment\'o en t\'erminos de publicaciones, comentarios y respuestas mucho antes de que comenzara la etapa oficial de campa\~nas. Para mayo de 2014, la tasa de contribuci\'on p\'ublica alcanz\'o niveles que se mantuvieron durante varios meses. Tambi\'en para esa fecha, el candidato publicaba contenido original a un ritmo que mantuvo hasta la toma de posesi\'on.

\begin{table}[]
\centering
\begin{tabular}{|l|l|c|c|c|}
\hline
\multicolumn{2}{|l|}{}                                                                                                    & \multicolumn{1}{l|}{Antes de la camapa\~na} & \multicolumn{1}{l|}{Durante la campa\~na} & \multicolumn{1}{l|}{Despu\'es de la camapa\~na} \\ \hline
\multirow{3}{*}{\begin{tabular}[c]{@{}l@{}}Promedio semanal \\ de contribuciones \\ del candidato\end{tabular}} & Publicaciones    & 7                                        & 12                                       & 6                                       \\ \cline{2-5} 
                                                                                                               & Comentarios & 7                                        & 9                                        & 16                                      \\ \cline{2-5} 
                                                                                                               & Respuestas  & 60                                       & 76                                       & 16                                      \\ \hline
\multirow{3}{*}{\begin{tabular}[c]{@{}l@{}}Promedio semanal \\ de contribuciones \\ de ciudadanos\end{tabular}}    & Publicaciones    & 19                                       & 240                                      & 137                                     \\ \cline{2-5} 
                                                                                                               & Comentarios & 71                                       & 1,100                                    & 156                                     \\ \cline{2-5} 
                                                                                                               & Respuestas  & 173                                      & 586                                      & 250                                     \\ \hline
\end{tabular}
\caption{Participaci\'on ciudadana en redes sociales antes, durante y despu\'es de la campa\~na}
\label{my-label}
\end{table}

Para resumir las tendencias en el uso de las redes sociales y la participaci\'on social, la Tabla 1 muestra los niveles de actividad promedio antes, durante y despu\'es de la etapa oficial de campa\~na. Tomando en cuenta el periodo previo a que \'esta comenzara, el candidato casi duplic\'o su ritmo de publicaci\'on en las redes sociales, con un promedio de 12 contribuciones por semana durante la campa\~na. Una vez que \'esta termin\'o, sus contribuciones en t\'erminos de publicaciones volvieron a los niveles iniciales. Llama la atenci\'on que la tasa de respuestas a las consultas de los usuarios cay\'o notablemente en comparaci\'on con los periodos antes y durante la campa\~na. Una vez que obtuvo el cargo, “El Bronco” particip\'o con la audiencia con comentarios y respuestas a los mensajes de los usuarios, m\'as que a trav\'es de sus propios mensajes. Tambi\'en notamos que la cantidad de publicaciones originales por semana regres\'o a niveles de precampa\~na, la tasa de comentarios sobre las publicaciones de los usuarios fue mayor que en cualquiera de los periodos anteriores, y la frecuencia de sus respuestas a los usuarios disminuy\'o notablemente. En general, una vez en el cargo, el uso de las redes sociales del candidato disminuy\'o en relaci\'on con el periodo activo de campa\~na.

La Tabla 1 tambi\'en revela una transici\'on en la forma en que la audiencia particip\'o con el candidato en Facebook. En los meses previos a la etapa oficial de campa\~na, los usuarios interactuaron con “El Bronco” principalmente a trav\'es de respuestas a los comentarios del candidato; durante la campa\~na, lo hicieron con nuevos comentarios m\'as que con respuestas. Una vez terminada la campa\~na, el ritmo de participaci\'on social se mantuvo alto; sin embargo, la distribuci\'on entre los diferentes modos de participaci\'on se hizo m\'as equilibrada: los usuarios generaron una proporci\'on m\'as balanceada de publicaciones, comentarios y respuestas.

Los usuarios tuvieron un gran n\'umero de conversaciones a trav\'es de las redes sociales, y durante el punto m\'as \'algido del periodo electoral, reaccionaron a los mensajes y las consultas del candidato con mil 100 comentarios a la semana. Lo m\'as interesante es que la participaci\'on p\'ublica en asuntos pol\'\i{}ticos a nivel estatal no se evapor\'o despu\'es de la etapa oficial de campa\~na. Siguiendo la hip\'otesis del slacktivismo podr\'\i{}amos esperar que la participaci\'on p\'ublica en esta plataforma disminuyera con el tiempo y declinara con rapidez tras el d\'\i{}a de la elecci\'on, cuando las conversaciones pol\'\i{}ticas suelen pasar a otros temas. Si bien el ritmo de participaci\'on en las redes sociales al finalizar la campa\~na disminuy\'o en comparaci\'on con los niveles mostrados durante \'esta, se mantuvo por encima del periodo anterior a la campa\~na. En otras palabras, la p\'agina de Facebook de “El Bronco” se convirti\'o en una plataforma sostenible de conversaci\'on p\'ublica, ya que los usuarios continuaron elaborando publicaciones, comentando y respondiendo con mucha mayor frecuencia que antes del inicio de la campa\~na. 

Resulta claro que hubo alg\'un tipo de retroalimentaci\'on positiva entre el uso de las redes sociales del candidato y la voluntad de los usuarios de participar sobre temas de pol\'\i{}tica en la plataforma de Facebook. El \'ultimo paso en este an\'alisis es estimar hasta qu\'e punto la cantidad de publicaciones, comentarios y respuestas de los usuarios impuls\'o el uso de las redes sociales de “El Bronco”. En ese mismo sentido, ?`qu\'e tipos de interacci\'on con el candidato motivaron a los usuarios a seguir participando m\'as all\'a del d\'\i{}a de la elecci\'on?

Para responder a esta pregunta, realizamos un an\'alisis de regresi\'on para elaborar un modelo del v\'\i{}nculo entre la interacci\'on pol\'\i{}tica directa de los usuarios con los pol\'\i{}ticos en redes sociales y la continuidad de su participaci\'on en asuntos pol\'\i{}ticos en la misma plataforma. As\'\i{}, realizamos un an\'alisis de regresi\'on log\'\i{}stica con medidas repetidas para observar si la interacci\'on directa con “El Bronco” predijo un compromiso posterior en su p\'agina de seguidores. Incluimos las medidas de participaci\'on de publicaciones en la p\'agina, comentarios a las publicaciones y respuestas a los comentarios en la p\'agina como variables independientes en el modelo. Nuestro an\'alisis indica que la participaci\'on del candidato por medio de respuestas individuales a los usuarios es un predictor significativo y positivo de la probabilidad de que dichos ciudadanos vuelvan a publicar, comentar y responder en la p\'agina ($\beta$= 6.66, $p$ \textless 0.00 para comentarios; $\beta$= 0.91, $p$ \textless 0.00 para respuestas). La relaci\'on inversa result\'o un poco m\'as d\'ebil: la participaci\'on individual de un ciudadano con “El Bronco” no aumenta la probabilidad de que \'este publique, comente o responda activamente a dicho ciudadano  ($\beta$= 0.47, $p$ \textless 0.00 para comentarios; $\beta$=0.08 para publicaciones and $\beta$= 0.14, $p$ \textless 0.002 para respuestas). Recibir comunicaci\'on directa y en l\'\i{}nea por parte de los ciudadanos rara vez motiva al candadito a responder. 

En general, estos resultados sugieren que las interacciones directas y personales con los pol\'\i{}ticos pueden impulsar a los ciudadanos a aumentar su participaci\'on en el di\'alogo p\'ublico sobre pol\'\i{}tica. Parece ser que cuanto m\'as interact\'ua un candidato con los ciudadanos, m\'as se involucran estos en pol\'\i{}tica en Internet. Esto es cierto para publicaciones de Facebook, comentarios y respuestas. Lo mismo ocurre con la forma en que el candidato participa con los ciudadanos: cuanto m\'as reciba respuestas y comentarios de los usuarios y m\'as publicaciones lo involucren, m\'as comentar\'a, responder\'a y compartir\'a con la comunidad.

\section{Conclusi\'on}
Este caso es una valiosa muestra del papel que las redes sociales pueden tener en diversas formas de la participaci\'on ciudadana. En esta elecci\'on estatal en M\'exico, un candidato desarroll\'o una estrategia de comunicaci\'on pol\'\i{}tica enfocada principalmente en las redes sociales. Dicho candidato gan\'o la elecci\'on y su p\'agina p\'ublica de Facebook se convirti\'o en una infraestructura central de informaci\'on que posibilit\'o el di\'alogo entre ciudadanos sobre pol\'\i{}ticas p\'ublicas y lo continu\'o haciendo tiempo despu\'es del d\'\i{}a de las elecciones. Este caso resulta significativo ya que es uno de los primeros ejemplos claros de un candidato pol\'\i{}tico que utiliza con \'exito las redes sociales para obtener un cargo de elecci\'on p\'ublica y establecer un plataforma de di\'alogo p\'ublico. Sin duda existen otros casos de campa\~nas fallidas y de intercambios en las redes sociales que no fueron civilizadas ni se mantuvieron por mucho tiempo. Sin embargo, nuestros hallazgos son consistentes con el creciente consenso de que los directores de campa\~nas pol\'\i{}ticas reconocen a las redes sociales como un medio para la participaci\'on de los candidatos con los votantes \cite{kreiss2016prototype,nielsen2012ground}.

?`Es cierto entonces que el uso de las redes sociales tiene consecuencias negativas en la participaci\'on ciudadana? Nuestros resultados muestran que en una elecci\'on en que tanto el candidato como los ciudadanos utilizan las redes sociales con comodidad, el efecto es positivo para ambos tipos de actores pol\'\i{}ticos. En primer lugar, los candidatos encuentran a sus partidarios y conocen a sus votantes; en segundo, los ciudadanos se involucran con los candidatos y esperan que la participaci\'on se mantenga incluso despu\'es del d\'\i{}a de la elecci\'on y de la toma de posesi\'on. Esto pone en cuestionamiento la hip\'otesis del \textit{slacktivismo} y su presunci\'on de que la participaci\'on ciudadana en las redes sociales no hace m\'as que promover una participaci\'on ef\'\i{}mera que no se refleja en el mundo fuera de las redes sociales \cite{christensen2011political,rotman2011slacktivism}.

Con demasiada frecuencia, diversas formas de activismo digital son desacreditadas y tildadas de \textit{slacktivismo}, lo cual clausura la investigaci\'on sobre las formas en que la gente combina el uso de medios y el di\'alogo pol\'\i{}tico en sus vidas diarias. La evaluaci\'on de la hip\'otesis del  \textit{slacktivismo} requiere que reconozcamos el rango de posibilidades de llevar a cabo acciones proporcionado por las redes sociales, as\'\i{} como las plataformas de medios y los patrones de uso que definen la cultura pol\'\i{}tica contempor\'anea \cite{nagy2015imagined}. El supuesto de que la participaci\'on ciudadana disminuye al aumentar la participaci\'on en redes sociales puede ser evaluado tanto en t\'erminos de calidad como de cantidad de interacciones p\'ublicas (en este art\'\i{}culo evaluamos dicho supuesto con evidencia sobre el volumen y la frecuencia de las interacciones). De manera anecd\'otica, observamos que la p\'agina del “El Bronco” comenz\'o como un foro para la discusi\'on y el posicionamiento pol\'\i{}tico antes del inicio de la campa\~na, y se transform\'o en una plataforma para que la gente presentara quejas y peticiones p\'ublicas al t\'ermino de \'esta. Investigaciones posteriores podr\'\i{}an examinar las caracter\'\i{}sticas de determinados formatos de publicaci\'on o la sofisticaci\'on pol\'\i{}tica y el prop\'osito de las contribuciones. 

Con base en la evidencia que recabamos aqu\'\i{}, si tanto l\'\i{}deres pol\'\i{}ticos como ciudadanos utilizan la misma plataforma de redes sociales, en el contexto de elecciones democr\'aticas, podemos esperar algunos resultados de participaci\'on positivos. Los datos revelan que la participaci\'on del candidato con los ciudadanos en la p\'agina de seguidores de Facebook tuvo un efecto positivo, dando como resultado un uso sostenido de la plataforma. Adem\'as, la participaci\'on ciudadana continua tuvo un efecto positivo en el candidato. Su ritmo de participaci\'on no se mantuvo uniforme durante el periodo completo de campa\~na, pero permaneci\'o notablemente alto incluso despu\'es de que asumi\'o el cargo y, probablemente, obtuvo un acceso m\'as directo a periodistas y medios de comunicaci\'on. 

Como siempre, la selecci\'on de casos y la metodolog\'\i{}a proporcionan tanto fortalezas como l\'\i{}mites a la generalizaci\'on. Existen varias formas de evaluar la hip\'otesis del \textit{slacktivismo}, y en este trabajo nosotros hacemos una aportaci\'on al investigar los cambios en la frecuencia de las contribuciones. En el futuro, ser\'a importante estudiar las cualidades del mensaje y el rango de factores informativos, sociales, pol\'\i{}ticos, psicol\'ogicos o tecnol\'ogicos que pueden moderar la relaci\'on entre el uso de las redes sociales y la participaci\'on ciudadana. Los candidatos a cargos de elecci\'on p\'ublica en otros pa\'\i{}ses —a trav\'es de otras plataformas de redes sociales— tendr\'an diferentes capacidades y limitaciones. M\'as a\'un, es posible imaginar una amplia gama de hip\'otesis condicionales. Si “El Bronco” hubiera perdido las elecciones y hubiese dejado la pol\'\i{}tica, ?`los ciudadanos que hab\'\i{}an sido atra\'\i{}dos al di\'alogo pol\'\i{}tico en Facebook seguir\'\i{}an participando activamente en temas pol\'\i{}ticos en la plataforma? ?`Habr\'\i{}an permanecido involucrados y s\'olo se habr\'\i{}an cambiado de p\'agina? Si bien es posible analizar el ritmo del di\'alogo pol\'\i{}tico, los datos en realidad no revelan qu\'e usuarios votaron ni por qu\'e candidato lo hicieron. Sin embargo, estas hip\'otesis y reservas no merman la tesis de que cuando los candidatos a puestos de elecci\'on p\'ublica y la poblaci\'on utilizan las redes sociales para el di\'alogo pol\'\i{}tico, tienen la capacidad de crear nuevos patrones de participaci\'on ciudadana que pueden durar varios meses m\'as all\'a del d\'\i{}a de las elecciones.


\newpage
\setcounter{page}{1}
\renewcommand{\thepage}{}
\renewcommand{\refname}{Bibliograf\'\i{}a}
\bibliography{proposal}

\newpage
\setcounter{page}{1}

\thispagestyle{empty}
\textbf{\huge Social Media, Civic Engagement, and the Slacktivism Hypothesis: Lessons from Mexico's ``El Bronco''
} \\ 
\\
\vspace{-0.3pc}
\emph{Philip N. Howard, Oxford University, Saiph Savage, Universidad Nacional Autonoma de Mexico (UNAM)/West Virginia University, Claudia Flores-Saviaga, West Virginia University, Carlos Toxtli, West Virginia University, Andrés Monroy-Hernández, Microsoft Research}
\vspace*{0.1in}

\vspace{0.4pc}

\begin{abstract}
Does social media use have a positive or negative impact on civic engagement? The cynical ``slacktivism hypothesis'' holds that if citizens use social media for political conversation, those conversations will be fleeting and vapid. Most attempts to answer this question involve public opinion data from the United States, so we offer an examination of an important case from Mexico, where an independent candidate used social media to communicate with the public and eschewed traditional media outlets. He won the race for state governor, defeating candidates from traditional parties and triggering sustained public engagement well beyond election day. In our investigation, we analyze over 750,000 posts, comments, and replies over three years of conversations on the public Facebook page of ``El Bronco.'' We analyze how rhythms of political communication between the candidate and users evolved over time and demonstrate that social media can be used to sustain a large quantity of civic exchanges about public life well beyond a particular political event.\footnote[1]{Published in: \textit{Journal Of International Affairs} Winter 2016, Vol. 70 Issue 1, p55-73. 19p. ISSN:0022-197X}
\end{abstract}

\newpage

\vspace{0.75pc}

\textbf{\huge Social Media, Civic Engagement, and the Slacktivism Hypothesis: Lessons from Mexico's ``El Bronco''
} \\ 
\\

\label{sec:technical}
\vspace{-0.3pc}
\section{Introduction}
Social media have become an important part of modern political campaigning. Campaign managers mine them for data. Citizens and civic groups use a plethora of platforms, such as Twitter, Facebook, and reddit, to talk about politics and engage with civil society groups and political leaders \cite{karpf2012moveon}. Candidates and political parties are also using social media to manage their public image in communications with journalists and the interested public \cite{gainous2014tweeting}. While many factors affect whether or not political actors adopt social media, the vast majority today are at least actively trying to integrate social media into their campaigns \cite{nah2012modeling}.  The problem is that despite the technological advancements, most political and activist groups are still in the dark on how best to mobilize people \cite{lovejoy2012information}. The main difficulty arises because citizens’ decisions about how much to participate in a political cause depend on how they perceive the efforts of the political candidate or organization \cite{savage2016botivist}. But the norms on social media platforms are continually shifting. As a result, people are constantly changing how they interpret the social media content and online efforts of others. It can thus be very difficult and time consuming for political groups to keep up with new technology and to predict the outcomes of sharing certain types of media and content. This complexity has forced many political organizations and candidates to limit how much they use social media and who among their organization or group can use social media \cite{obar2014canadian}. Politicians usually prefer to have a point person in charge of their party’s social media strategy and even their own personal accounts. However, even when the point person finally understands how to mobilize citizens, it is not easy to transfer that knowledge to others.  As a result, most politicians are very cautious about how they use social media and how much they interact with their online audiences \cite{savage2015participatory}. This has hindered and limited our understanding of which political strategies work best to mobilize and engage with citizens.

Our understanding of social media and elections has also been bounded by the fact that most of the research has focused on the United States. In this relatively advanced democracy, social media has become a tool for some forms of civic engagement and political expression.  Among the advanced democracies, this seems to have resulted in only modest forms of activism, such as petition signing or sharing political content from affinity groups over networks of family and friends \cite{gibson2000social}.  Therefore, research on social media and civic engagement in U.S. politics has often sought to test, directly or indirectly, the ``slacktivism hypothesis.''  We define the slacktivism hypothesis as the supposition that if Internet or social media use increases, civic engagement declines. 

The argument that the use of social media has mostly negative consequences for public life begins with evidence that most of the content shared over social media is rarely about politics, and even when it is about politics it consists of short messages shared by people with short tempers in short conversations  \cite{katz2013social}. Citizens rarely use social media for substantive political conversations, and such conversations are often anemic, uncivil, or polarizing.  Overall, online political conversations are relatively rare occurrences in comparison to the other kinds of things people do on the Internet on a daily basis  \cite{massanari2011information}. When they do occur, moreover, during major political events such as candidate debates, social media users will use digital platforms to learn about and interact with politics, but they tend to acquire new knowledge that is favorable to their preferred candidate \cite{boulianne2015social}. Recent work has found that while many U.S.-based activist organizations believe that they are creating stronger communities and dialogues with their public through social media content, this rarely translates to significant mobilization with regard to public events, consumer activism, or grassroots lobbying \cite{guo2013tweeting,lovejoy2012information}.

Researchers have demonstrated that social media use causes people to turn their social networks into ``filter bubbles'' that diminish the chance of exposure to new or challenging ideas. In other words, social media allow us to create homophilous networks \cite{pariser2011filter}. For example, massive amounts of Twitter data have been used to classify users by party affiliation and homophily in the United States, with results indicating that average Democrats tend to be more homophilous than average Republicans, unless the users classified as Republicans follow major Republican leaders \cite{colleoni2014echo}. Ultimately, public debates over social media may do little more than promote ephemeral engagement without translating to offline political impact \cite{christensen2011political,rotman2011slacktivism}.  When social media actions do have offline impacts, they are usually the same kinds of low-quality, high-volume actions that advocacy and political groups have long used to gain notoriety and news headlines for their organizations \cite{karpf2010online}.

In the early stages of the slacktivism debate, there appeared to be only a few 
very specific cases of movements originating from the Internet that both successfully mobilized people and achieved public policy goals, and often these cases involved a narrow range of technology access issues \cite{benkler2015social,faris2015score,freedman2016strategies, dubois2012fifth,kreiss2016prototype,nielsen2012ground}.  In more recent years the distinction between online and offline political action has evaporated, such that modern political candidates need to be savvy with multiple technology platforms and many kinds of campaigns spend significant resources on data analytics.  There are now multiple examples of traditional social movements that have scored impressive victories through their effective use of social media, as well as new social movements that have originated online and become stable civil society actors \cite{beyer2014expect}.  And complicating all of this is the growing problem of algorithmic control over social media messaging:  automated programs can be used to activate citizens or to discourage their engagement \cite{savage2015participatory,woolley2016automation}. 

The argument against the slacktivism hypothesis is that political engagement over social media is always in addition to, not a replacement for, whatever citizens would normally be doing in their political lives \cite{christensen2011political}. There are important public conversations occurring over social media that grow especially intense during important political events.  For example, research has found that social media use helps people build their political identity and community awareness, which even results in financial contributions to relevant civil society groups \cite{lee2013does}. Indeed, social media, like other Internet-based communications, tend to supplement our intake of information about politics, elections, and public policy, and allow people to be more omnivorous in their information diets. Such ``political omnivores'' still rely on major broadcast media for information but regularly depend on the Internet for interactivity about politics \cite{massanari2011information}.

There is evidence that young adolescents’ use of social media—in conjunction with the intent to participate and the consumption of television news—creates a virtuous circle of civic engagement \cite{kruikemeier2016news}. Most research has consequently focused primarily on small-scale surveys or interview studies. Additionally, it has proven difficult to actually measure levels of civic engagement over social media platforms, especially because there is usually not one central social media site that citizens and politicians use. Being exposed to online activism might influence individual decisions on subsequent civic actions, such as signing a petition or donating to charity, but it is not clear that the use of social media for political conversation results in more sophisticated voters or an increased probability of voter turnout. 

While researchers debate the slacktivism hypothesis in the context of the United States and advanced democracies, there are good reasons to expect the relationship between media use and civic engagement in international contexts to be different. The first reason is straightforward for scholars of international studies: there is such a great variety of regime types and political institutions around the world that we should not expect evidence from the United States to hold in many other contexts. The second is a more specific observation from comparative media systems research: the world outside the United States produces, consumes, and regulates political news and information in very different ways \cite{wessler2016global}.

First, most of the world thinks the Internet is Facebook, and a significant amount of the time many users spend online is actually spent on the Facebook platform \cite{stewart2016facebook}. The Internet that activists, citizens, and voters use to consume political information in the United States and many advanced democracies is experienced through a browser on a personal computer, and increasingly on a smart phone. In regions of the world where data plans are more expensive and bandwidth more erratic, people instead use SMS (text messaging) or the built-in applications that come with less expensive cellphones. As a result, Facebook has become synonymous with the Internet for most people around the world.

However, the vast majority of the research on slacktivism has been conducted in the United States, and international case studies on the topic demonstrate, sensibly, that the relationship between social media diffusion and civic engagement is complex. Some scholars have demonstrated that political leaders rarely use the most interactive platforms for fear of losing control of the content they produce and the messages they craft \cite{stromer2014presidential,howard2003digitizing}. Morozov has argued that many digital media platforms are unable to sustain the attention of people who offer a few clicks of support through online petitions but have little energy for the kinds of political engagement that take time or involve personal risk \cite{morozov2014save}. He draws from several international examples of social movements that may have failed because of their dependence on information technologies, but his is not the most systematic analysis.

The literature on international communication is vast, but there are lessons to draw from scholars tackling the study of contemporary political communication. Close study of authoritarian regimes, from Azerbaijan to China, has revealed that social media can be an important means of conducting political conversations, but that doing so depends on the censorship interests and fabrication activities of ruling elites \cite{king2016chinese,king2013censorship,pearce2015affordances}. Bailard has demonstrated through a natural experiment that digital media use during elections in Tanzania and Bosnia resulted in increased levels of civic engagement, negligible impact on voter turnout, and higher rates of voter cynicism \cite{anderson2016democracy,bailard2012field}. Moreover, social media use is an important part of the causal explanation for the shape and character of a growing number of major international public protests, including those in Chile and Hong Kong, where Facebook and Twitter use for news proved to be important predictors of protest participation, even holding constant factors like post-materialist values and political ideology \cite{valenzuela2012social,lee2015social}. A long-term study of trust in institutions in seven countries across Asia has revealed that social media are particularly important in raising civic engagement in countries where other media options are meager \cite{lee2016digital}. An early analysis of four advanced democracies revealed that fears that the Internet has a broad negative effect on social capital and political participation were unfounded \cite{gibson2000social}.

The bloom of pro-democracy protests may also undermine the slacktivism hypothesis. Social movements are always causally complex phenomena. However, activists and protest leaders say social media were essential to the organization of the protests; government leaders try to respond over social media or censor it because they believe it to be integral to the protest; journalists report on the particular dynamics of social media use in their coverage of events; and scholars, in hindsight, find it difficult to develop an analytical narrative about events without discussing the role of social media \cite{monroy2015narcotweets}.

\section{The Case of ``El Bronco''}
Finding international cases that help us test the slacktivism hypothesis is a challenge because there are few political leaders, in either authoritarian or democratic regimes, whose careers have been built on the savvy use of social media.  The evidence on social media and slacktivism is also encumbered by the fact that in every one of the media systems discussed, social media are only a small part of the communications strategy for political leaders. Until recently, social media campaigning was mostly about reaching journalists and other policy makers, rather than a broad public \cite{kreiss2014seizing}.  Most media systems have a dominant means by which citizens get news and information about politics, and usually it is television, radio, or newspapers. The Internet, and social media in particular, provides a secondary source of media, though interestingly it is—among all the possibilities—the most commonly chosen secondary medium \cite{massanari2011information}. In other words, citizens often get most of their news and information either from the television, radio, or newspaper, and then check sources, poll their friends and family, or do additional research on the Internet.

This too may be changing. In 2015, Jaime Rodriguez Calderon, a Mexican independent candidate also known as ``El Bronco,'' was elected governor in the Mexican state of Nuevo Leon on the basis of a campaign that treated traditional media with disdain and actively engaged with the electorate through Facebook and Twitter \cite{broncoMex}.  Unlike most candidates, ``El Bronco'' did not pay for TV ads.  Even street advertisements, very common in Mexican elections, were done by community supporters rather than through a centralized campaign organization.  In this paper, we analyze the entirety of online interactions between citizens and ``El Bronco,'' who, primarily through Facebook, was able to run as an independent and defeat his closest rival by nearly 25 percentage points \cite{broncoMex2}. 
Nuevo Leon is a northern state in Mexico with around five million inhabitants, the second highest development scores in the country, and the second highest rates of Internet penetration in the country. Close to 60 percent of households have access to the Internet, comparable to states like Mississippi in the United States \cite{broncoMex3,broncoMex4,rainie2014census}. 

The case of ``El Bronco'' is valuable precisely because it is an extreme case of social media use during a competitive election. It highlights the notable outcome of an independent candidate winning an election through dedicated social media engagement with voters. This major subnational election provides a unique opportunity to answer an important research question:  what is the impact of social media use on civic engagement during the campaign season, and beyond election day itself?

\section{Data Collection and Analysis}
\begin{figure}
\centering
  \includegraphics[width=1.0\columnwidth]{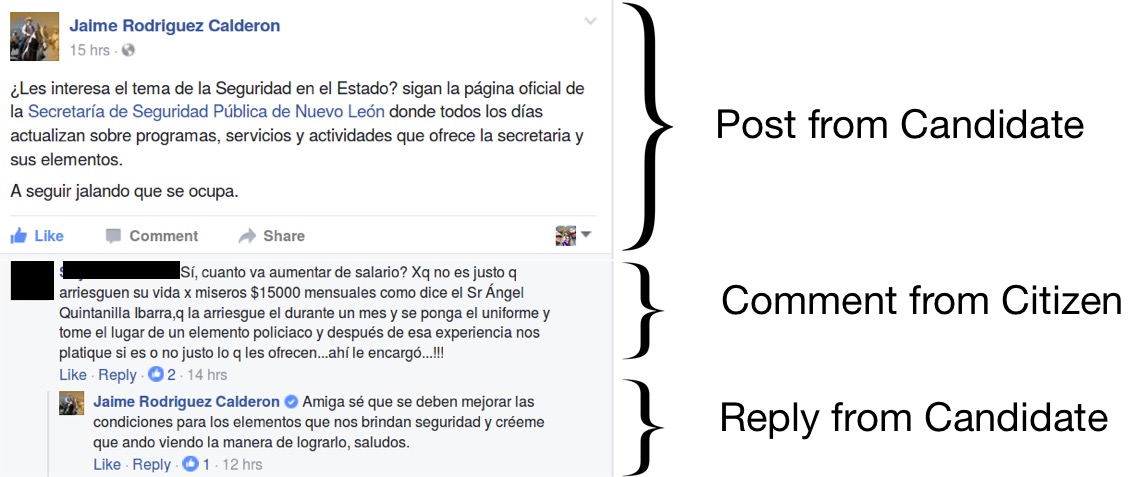}
  \caption{Conversation initiated by Jaime Rodriguez Calderon.  }\label{fig:postFacebook}
\end{figure}
To study the long-term patterns of political engagement over social media, we collected three years of posts, comments, and replies from the Facebook fan page of Jaime Heliodoro Rodriguez Calderon\footnote{\url{https://www.facebook.com/JaimeRodriguezElBronco}}. Using the platform's application program interface (API) we collected 5,708 posts from the candidate and 71,446 from citizens; 20,045 comments from the candidate and 458,544 from citizens; and 31,527 comment replies from the candidate and 171,577 from citizens. This method captured all activity between November 2012, when Bronco’s social media presence was created, and April 2016, almost one year after he took office.   

Our analysis began with the collection of many kinds of data, from many people, generated over many kinds of devices.  We created a series of online bots to query the Facebook API and collect data on posts, images, comments, and likes from Bronco’s fan page. We used MongoDB, a non-relational database, to help manage the process of saving large volumes of data generated by our queries and to help reconcile the changes over the years in how Facebook organizes data about posts and user devices.

These data capture political conversations initiated either by the candidate or by a Facebook user. They consist of posts on the politician’s fan page, user comments, and subsequent replies. To help explain the structure of this data, Figure 1 presents the organization of a typical conversation initiated by the candidate. The candidates makes a post that attracts a comment from a citizen. The candidate may then comment on what the citizen has written, and subsequently there can be quite a long thread of discussion through replies. Figure 2 presents the organization of a political conversation initiated by a citizen. The citizen makes a post on the politician’s page and possibly triggers engagement when the candidate comments on the post. The citizen may then reply and trigger an extended exchange with the candidate. The posts, comments, and replies can be viewed by any user, and other users can subsequently generate their own posts.
\begin{figure}
\centering
  \includegraphics[width=.6\columnwidth]{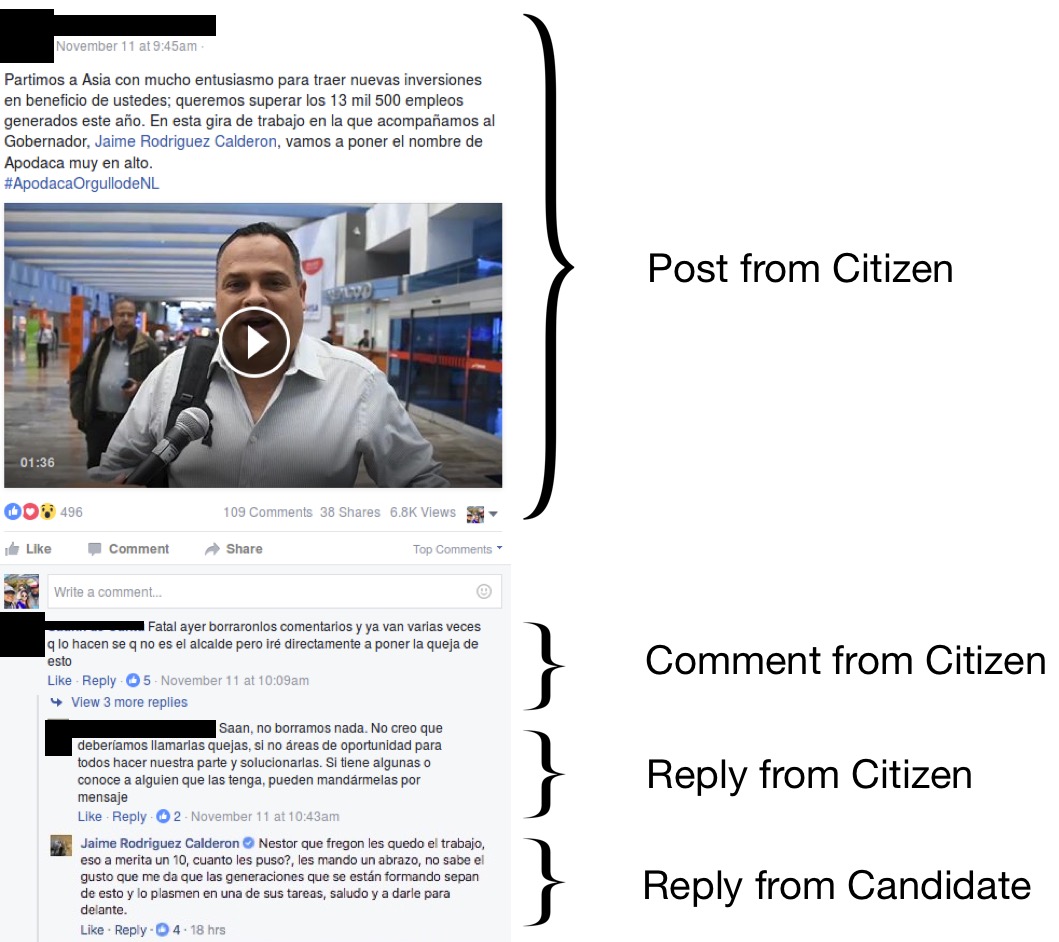}
  \caption{Conversation started by citizen. }\label{fig:postFacebook2}
\end{figure}
To study the communication patterns of the candidate and citizens, we plotted the moving averages of numbers of posts, comments, and replies. We use a 15-day statistical moving average to allow us to display a smooth, more interpretable trend line over the three-year period.

``El Bronco'' became an independent politician when he left the Institutional Revolutionary Party (PRI) and, in November 2012, created a Facebook fan page from which to develop his new political profile. At the time, he claimed that his break from the PRI and disavowal of major broadcast media in favor of social media were of themselves important markers of his new public identity. Through much of 2013 the amount of traffic on his social media platform was limited, but by 2014 his hardline stance against organized crime and government corruption found a growing audience. Moreover, he was clearly gaining popularity many months before the election, and by June 2014 he was writing multiple posts, comments, and replies each day, a pace of social media engagement that he was able to sustain for the following two years.
\begin{figure}
\centering
  \includegraphics[width=.8\columnwidth]{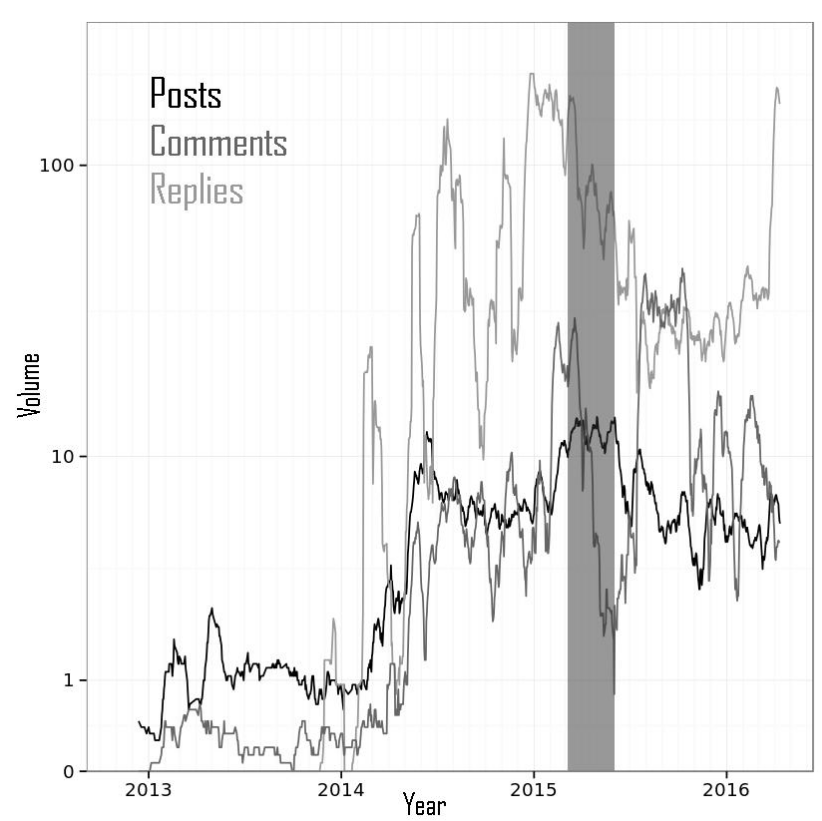}
  \caption{Rhythm of candidate-authored social media content }\label{fig:postFacebook}
\end{figure}

\begin{figure}
\centering
  \includegraphics[width=.8\columnwidth]{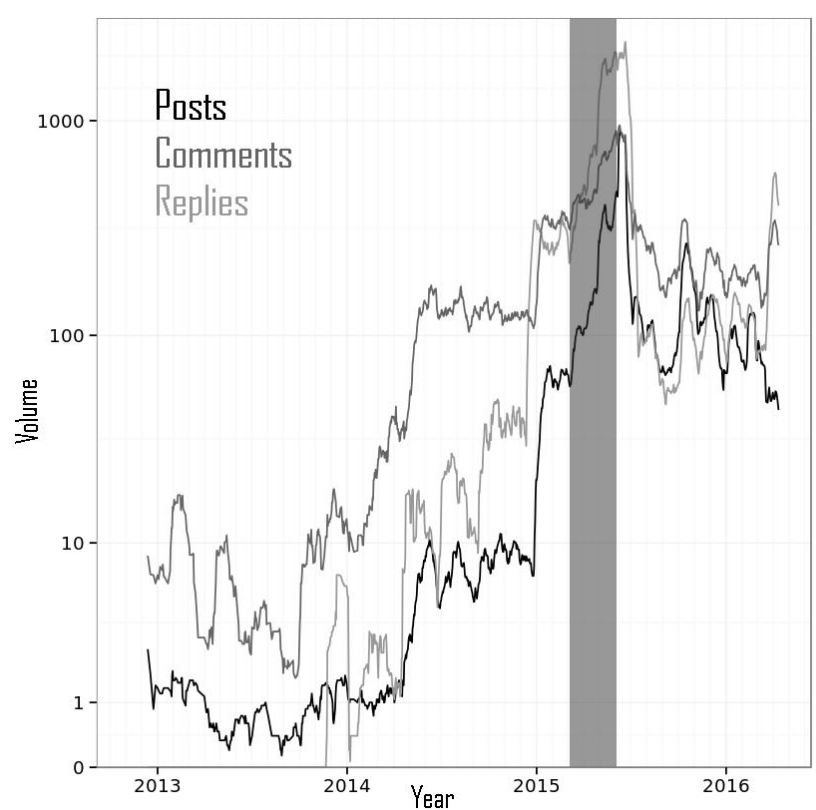}
  \caption{Candidate's Facebook  fan page }\label{fig:postFacebook}
\end{figure}

Figures 3 and 4 plot the moving average number of posts, comments, and replies over the entire period. Even though the official campaign period for the office of state governor only lasted three months, it is clear that civic engagement on the fan page peaked well beforehand. Indeed, the analytical narrative for this data should have three chapters: the pre-campaign period November 2011 – February 2015; the official campaign period March – May 2015; and the time in elected office from June 2015 to the present. In these figures, many of the peaks in political conversation over social media occurred during big moments in Mexican and international politics, and many of the valleys coincide with public holidays. In Figures 3 and 4, the dark gray band indicates the beginning and end of the official gubernatorial campaign, which ran from March 6 to June 3, 2015.

Figure 4 reveals that the pace of public conversation, both around politics and the candidate, was increasing in terms of posts, comments, and replies well before the formal campaign period began. By May 2014, the public rates of contribution had already hit levels that were sustained for many months thereafter. By that date, the candidate was also posting original content at a pace that he sustained right through to taking office.

To further summarize the trends in social media use and civic engagement, Table 1 presents the weekly average levels of activity before, during, and after the official campaign period. In the lead-up to the campaign, the candidate doubled his pace of posting to social media, making 12 contributions per week, on average, during the campaign. After the campaign, his contributions (in terms of posts) returned to his pre-campaign rate. Interestingly, the rate of replies offered to citizen queries dropped off noticeably compared to both the campaign period and before the campaign period. Once in office, El Bronco tended to engage the public with comments and replies to user posts more than through his own original posts. The number of original posts each week returned to the pre-campaign level, his rate of commenting on user posts was higher than in either of the two previous periods, and the frequency of his replies to users diminished very noticeably. Overall, once in office, the candidate’s social media use diminished relative to that of his active period of campaigning.  

\begin{table}[]
\centering
\begin{tabular}{|l|l|c|c|c|}
\hline
\multicolumn{2}{|l|}{}                                                                                                    & \multicolumn{1}{l|}{Before the campaign} & \multicolumn{1}{l|}{During the campaign} & \multicolumn{1}{l|}{After the campaign} \\ \hline
\multirow{3}{*}{\begin{tabular}[c]{@{}l@{}}Weekly average \\ contributions from \\ the candidate\end{tabular}} & Posts    & 7                                        & 12                                       & 6                                       \\ \cline{2-5} 
                                                                                                               & Comments & 7                                        & 9                                        & 16                                      \\ \cline{2-5} 
                                                                                                               & Replies  & 60                                       & 76                                       & 16                                      \\ \hline
\multirow{3}{*}{\begin{tabular}[c]{@{}l@{}}Weekly average \\ contributions\\ from \\ citizens\end{tabular}}    & Posts    & 19                                       & 240                                      & 137                                     \\ \cline{2-5} 
                                                                                                               & Comments & 71                                       & 1,100                                    & 156                                     \\ \cline{2-5} 
                                                                                                               & Replies  & 173                                      & 586                                      & 250                                     \\ \hline
\end{tabular}
\caption{Civic engagement over social media before, during, and after the election campaign}
\label{my-label}
\end{table}

Table 1 also reveals a transition in the way the public engaged with the candidate on Facebook. In the months before the formal campaign period, users engaged with “El Bronco” primarily through replies to each other’s comments.  During the campaign, users primarily engaged through fresh new comments rather than replies. After the campaign, the pace of social engagement remained high but the distribution among different modes of engagement became more balanced: citizens generated a more equal ratio of posts, comments, and replies. 

Users drove large numbers of conversations over social media, and during the peak of the election period they were reacting to the candidate’s messages and queries with 1,100 comments a week. Most interestingly, public engagement with state-level political issues did not evaporate after the official campaign period. If the slacktivism hypothesis were true, we might expect public engagement on this platform to decline over time, and to decline quickly after election day when public conversations move on to other topics.  While the pace of social media engagement after the campaign diminished in comparison to that of the campaign period, it remained much higher than that of the period before the campaign. In other words, El Bronco’s page became a sustainable mode of public conversation as users continued to post, comment, and reply with much more frequency than before the campaign.

Clearly there was some kind of positive feedback between the candidate’s energetic social media use and users’ willingness to engage in politics on the Facebook platform. The final step in this analysis is to estimate the degree to which the number of posts, comments, and replies from users drove El Bronco’s use of social media. Similarly, what types of interaction with the candidate drive users to stay engaged beyond election day? 

To answer this question, we conducted a regression analysis to model the relationship between citizens’ direct interaction with politicians over social media and their continued engagement with political issues on the same platform.  To do so, we conducted a repeated measure logistic regression analysis to see whether direct interaction with El Bronco predicted subsequent engagement on his fan page. We included the participation measures of posting on the page, commenting on a post, and replying to a comment on the page as independent variables in the model. Our analysis indicates that the candidate’s engagement with and replies to individual citizens is a significant, positive predictor of how likely those citizens will go on to post, comment, and reply on the page ($\beta$= 6.66, $p$ \textless 0.00 for comments; $\beta$= 0.91, $p$ \textless 0.00 for replies). The inverse relationship is a little weaker, and a particular citizen’s engagement with El Bronco does not raise the likelihood that he will post, comment, or reply in an engaging way with that citizen ($\beta$= 0.47, $p$ \textless 0.00 for comments; $\beta$=0.08 for posts and $\beta$= 0.14, $p$ \textless 0.002 for replies). Receiving direct online communication from citizens rarely motivates a candidate to reply.  

Overall, these results suggest that direct and personal interactions with politicians can lead citizens to participate more in public conversations about politics and policy.  It appears that the more that the candidate directly interacts with citizens, the more citizens participate in politics online, and this is true for Facebook posts, comments, and replies. The same is true for how the candidate engages with citizens: the more the candidate receives replies and comments from citizens, or posts that involve them, the more the candidate will comment, reply to, and share with the community.

\section{Conclusion}
This case is a valuable source of evidence of the role social media can have in extended forms of civic engagement. In this Mexican state election, a candidate developed a political communication strategy focused primarily on social media. He won, and his public Facebook page became the key information infrastructure supporting public policy conversations among citizens well beyond election day.  This case is significant because it is one of the first clear cases of a political candidate successfully using social media to win elected office and sustain public conversation.  There are certainly other cases of failed campaigns and of social media conversations that were neither civil nor sustained.  But our findings are consistent with the growing research consensus that political campaign managers themselves see social media as a way for candidates to engage with voters \cite{kreiss2016prototype,nielsen2012ground}.

Does social media use have negative consequences for civic engagement? It turns out that in an election where a political candidate and citizens are comfortable using social media, the impact is positive for both kinds of political actors. Candidates find their supporters and learn about their constituents; citizens engage with candidates and come to expect engagement even after election day when candidates take office. This challenges the slacktivism hypothesis, which holds that civic engagement over social media does little more than promote ephemeral engagement and does not translate to the offline world \cite{christensen2011political,rotman2011slacktivism}.

Too often various forms of digital activism are dismissed as slacktivism, closing off inquiry into the ways in which people blend media use and political conversation into their daily lives.  Evaluating the slacktivism hypothesis requires that we appreciate the range of social media affordances, media platforms, and usage patterns that define contemporary political culture \cite{nagy2015imagined}.  The supposition that if social media increases civic engagement decreases can certainly be evaluated both in terms of the quality and quantity of public interactions—in this article we evaluate the hypothesis with evidence about the volume and frequency of interactions.  Anecdotally, we observed that the page began as a forum for political positioning and argument before the campaign and transitioned to a platform for people to submit public grievances and requests after the campaign.  Further research could examine the qualities of particular posting formats or the political sophistication and purpose of contributions.  

Based on the evidence collected here, if both political leaders and citizens use the same social media platform, in the context a democratic election, we may expect some positive engagement outcomes.  The data reveal that candidate engagement with citizens on the Facebook fan page had a positive effect, resulting in continued platform use. Moreover, continued citizen participation had a positive impact on the politician. His rhythm of engagement was not evenly sustained throughout the campaign period, but it remained noticeably high even after he took office and presumably had more direct access to journalists and broadcast media.

As always, case selection and methodology provide both strengths and limits to generalization. There are many ways to evaluate the slacktivism hypothesis, and here we take a step by investigating changes in the frequency of contributions.  In the future, it will be important to study message qualities and the range of other informational, social, political, psychological, or technological factors that may moderate the relationship between social media use and civic engagement.  Candidates for elected office in other countries, using other social media platforms, will face different capacities and constraints. Moreover, a wide range of counterfactuals can be imagined. If “El Bronco” had lost the election and left politics, would the citizens who had been drawn into political conversations on Facebook continue to actively engage in politics on that platform? Would they have remained engaged but moved off to another page? While it is possible to analyze the rhythm of political conversation, the data do not reveal which users actually voted or who they voted for. But these counterfactuals and caveats do not undermine the argument that when candidates for elected office and the public use social media for political conversation, they can create new patterns of civic engagement that can last for months beyond an election.

\newpage

\setcounter{page}{1}

\newpage
\setcounter{page}{1}




\newpage
\setcounter{page}{1}

\newpage
\setcounter{page}{1}


\newpage

\renewcommand{\thepage}{}

\renewcommand{\refname}{References}
\bibliography{proposal}

\end{document}